\documentclass[conference]{IEEEtran}
\pdfoutput=1

\usepackage{amsmath,amsfonts, bm}
\DeclareMathOperator*{\argmax}{arg\,max}
\DeclareMathOperator*{\argmin}{arg\,min}
\renewcommand{\vec}[1]{\mathbf{#1}}

\usepackage{cite}
\usepackage{amsmath,amssymb,amsfonts}
\usepackage{graphicx}
\usepackage{textcomp}
\usepackage{xcolor}
\usepackage[utf8]{inputenc}	
\usepackage[T1]{fontenc}		
\usepackage[hyphens]{url}
\usepackage{hyperref}            
\hypersetup{colorlinks=true,citecolor=blue,linkcolor=blue,urlcolor=blue,breaklinks=true}
\usepackage{url}				
\usepackage{booktabs}		
\usepackage{nicefrac}			
\usepackage{microtype}		
\usepackage{xspace}
\usepackage{multirow}
\usepackage{subcaption}
\usepackage{makecell}
\usepackage{algorithm}
\usepackage{algpseudocode}

\def\BibTeX{{\rm B\kern-.05em{\sc i\kern-.025em b}\kern-.08em
    T\kern-.1667em\lower.7ex\hbox{E}\kern-.125emX}}
\makeatletter

\def\ps@IEEEtitlepagestyle{
  \def\@oddfoot{\mycopyrightnotice}
  \def\@evenfoot{}
}
\def\mycopyrightnotice{
  {\begin{minipage}{\textwidth}\centering\footnotesize \copyright 2022 IEEE. Personal use of this material is permitted. Permission from IEEE must be obtained for all other uses, in any current or future media, including reprinting/republishing this material for advertising or promotional purposes, creating new collective works, for resale or redistribution to servers or lists, or reuse of any copyrighted component of this work in other works.\end{minipage}}
  \gdef\mycopyrightnotice{}
}

\begin{document}

\title{Search-based Methods for Multi-Cloud Configuration}


\author{\IEEEauthorblockN{1\textsuperscript{st} Malgorzata Lazuka}
\IEEEauthorblockA{\textit{IBM Research}\\
Zurich, Switzerland \\
\texttt{mal@zurich.ibm.com}}
\and
\IEEEauthorblockN{2\textsuperscript{nd} Thomas Parnell}
\IEEEauthorblockA{\textit{IBM Research}\\
Zurich, Switzerland \\
\texttt{tpa@zurich.ibm.com}}
\and
\IEEEauthorblockN{3\textsuperscript{rd} Andreea Anghel}
\IEEEauthorblockA{\textit{IBM Research}\\
Zurich, Switzerland \\
\texttt{aan@zurich.ibm.com}}
\and
\IEEEauthorblockN{4\textsuperscript{th} Haralampos Pozidis}
\IEEEauthorblockA{\textit{IBM Research}\\
Zurich, Switzerland \\
\texttt{hap@zurich.ibm.com}}}

\newcommand{\namedata}{MOCCA\xspace}
\newcommand{\namealgo}{CloudBandit\xspace}
\newcommand{\shortalgo}{CB\xspace}
\newcommand{\tabitem}{~~\llap{\textbullet}~~}

\maketitle

\begin{abstract}

Multi-cloud computing has become increasingly popular with enterprises looking to avoid vendor lock-in.
While most cloud providers offer similar functionality, they may differ significantly in terms of performance and/or cost.
A customer looking to benefit from such differences will naturally want to solve the multi-cloud configuration problem: given a workload, which cloud provider should be chosen and how should its nodes be configured in order to minimize runtime or cost?
In this work, we consider possible solutions to this multi-cloud optimization problem.
We develop and evaluate possible adaptations of state-of-the-art cloud configuration solutions to the multi-cloud domain.
Furthermore, we identify an analogy between multi-cloud configuration and the selection-configuration problems that are commonly studied in the automated machine learning (AutoML) field.
Inspired by this connection, we utilize popular optimizers from AutoML to solve multi-cloud configuration.
Finally, we propose a new algorithm for solving multi-cloud configuration, \namealgo{} (\shortalgo{}). 
It treats the outer problem of cloud provider selection as a best-arm identification problem, in which each arm pull corresponds to running an arbitrary black-box optimizer on the inner problem of node configuration.
Our extensive experiments indicate that 
(a) many state-of-the-art cloud configuration solutions can be adapted to multi-cloud, with best results obtained for adaptations which utilize the hierarchical structure of the multi-cloud configuration domain,
(b) hierarchical methods from AutoML can be used for the multi-cloud configuration task and can outperform state-of-the-art cloud configuration solutions and
(c) \shortalgo{} achieves competitive or lower regret relative to other tested algorithms, whilst also identifying configurations that have 65\% lower median cost and 20\% lower median time in production, compared to choosing a random provider and configuration.

\end{abstract}




\section{Introduction}

We live in the era of cloud computing, in which cloud providers compete to offer computing resources (servers, storage, managed software) to businesses and consumers via the internet. 
Cloud is a high-growth segment: according to \cite{Gartner}, worldwide end-user spending on public cloud services is forecast to grow 16.2\% in 2022 to total \$474 billion.
For enterprises, there are many benefits to using cloud services, e.g. increased flexibility, reduced costs, and rapid scaling. 

Coupled to the rise of cloud is the growing popularity of containers and container orchestration platforms.
Containers can be deployed quickly from one computing environment to another, making them well-suited to the world of cloud computing.
Many of today's workloads, such as machine learning (ML) training, are distributed in nature and require the deployment and management of multiple containers, all working together in parallel. 
Container orchestration platforms such as Kubernetes \cite{Kubernetes} enable running such systems resiliently, providing features like auto-scaling and failure management.

Most cloud providers offer a Kubernetes service 
that a user can utilize to specify the desired number of nodes and the nodes' configuration (how many CPUs, memory, etc.). 
Given a Kubernetes cluster, one can run a complex distributed workload, whilst remaining agnostic to which cloud provider is providing the hardware resources. 
This level of abstraction presents the user with a new optimization opportunity, which we will refer to as the \textit{multi-cloud configuration problem}. 
Namely, given a workload, which cloud provider should be selected, and how should the nodes in the cluster be configured, in order to minimize runtime or cost?

In practice, this optimization problem is of particular interest to users with complex, distributed workloads that must be run repeatedly. 
Consider a business with an ML-based recommendation engine that must be trained, in a distributed fashion, once every hour. 
By solving the multi-cloud configuration problem for such a recurring workload, the business will be able to exploit price/performance differences that may exist between cloud providers and save time or money with every run of the workload. 
However, any optimization algorithm will involve making a certain number of evaluations (e.g.,\ running the workload on a particular cloud provider, with a particular node configuration), and these evaluations themselves incur a certain additional cost.
In order for the savings offered by the optimized deployment to outweigh the expense of extra evaluations, effective algorithms with fast convergence are of particular interest in this context, and are the primary focus of this paper.

The existing solutions to solve the cloud configuration problem can be broadly separated into two categories: predictive methods and search-based methods. 
We provide a detailed overview of their pros and cons in Section~\ref{section:background}. 
In the first part of this work, we evaluate the performance of multiple such state-of-the-art methods, most of them adapted for the first time to the multi-cloud setting. In particular, we are interested in understanding 1) whether search-based methods are superior to predictive methods, and 2) how search-based methods developed in the single-cloud context should be adapted to the multi-cloud setting.
%
Furthermore, motivated by an interesting connection between the multi-cloud configuration problem and the selection/configuration problems studied in the automated machine learning (AutoML) field, we address yet another novel research question, namely how AutoML solutions perform when applied to the multi-cloud setting. Finally we propose a new search-based algorithm that outperforms the existing methods in terms of both convergence speed and cost savings. In summary, the contributions of our work are 4-fold:
\begin{itemize}
    \item As a result of a detailed review of the state-of-the-art, we have identified a set of promising predictive/search-based methods that we have implemented and adapted (when necessary) to multi-cloud configuration.
    \item We have applied (to our knowledge, for the first time) three popular AutoML methods to the multi-cloud configuration problem.
    \item Inspired by existing multi-armed bandit methods, we propose a novel, faster and more efficient algorithm for multi-cloud configuration, CloudBandit.
    \item We perform an extensive performance evaluation of CloudBandit, as well as of various predictive/search-based methods and AutoML solutions using a one-of-a-kind dataset with real cloud data collected from from 60 multi-cloud configuration tasks, across 3 public cloud providers.
    
    
\end{itemize}

%
%
%
%
%
%
%


\section{State-of-the-Art}
\label{section:background}

In this section, we will review the state-of-the-art in the field of predictive and search-based cloud configuration methods.
We will also discuss methods that try to solve an analogous optimization problem in the cloud scheduling literature.

\begin{table*}[]
\centering
\caption{State-of-the-Art Summary}
\label{tab:soa}
\begin{tabular}{lllccccl}
\hline
\textbf{Paper}        & \textbf{Type} & \textbf{Algorithms}                      & \multicolumn{1}{l}{\textbf{Offline}} & \multicolumn{1}{l}{\textbf{Online}} & \multicolumn{1}{l}{\textbf{Low-level Info.}} & \textbf{Multi-cloud} & \textbf{Baselines}                                                         \\ \hline
\cite{194946}         & Predictive    & Linear Regression (\textit{Ernest}) & $\times$                             & \checkmark                          & $\times$                             & $\times$             & n/a                                                                        \\
\cite{MARIANI2018618} & Predictive    & Random Forest                         & \checkmark                           & $\times$                            & \checkmark                           & $\times$             & n/a                                                                        \\
\cite{Yadwadkar}      & Predictive    & Random Forest (\textit{PARIS})                       & \checkmark                           & \checkmark                          & \checkmark                           & \checkmark           & \cite{194946}                                                              \\
\cite{216005}         & Predictive    & Collaborative Filtering (\textit{Selecta})              & \checkmark                           & \checkmark                          & $\times$                             & $\times$             & \cite{Yadwadkar}                                                               \\
\cite{201567}         & Search        & Bayesian Opt. (\textit{CherryPick})                & $\times$                             & \checkmark                          & $\times$                             & $\times$             & RS, CD, \cite{194946}                                                      \\
\cite{Bilal}          & Search        & Bayesian Opt., SHC, SA, TPE  & $\times$                             & \checkmark                          & $\times$                             & $\times$             & \cite{201567}                                           \\
\cite{8416333}        & Search        & Augmented Bayesian Opt. (\textit{Arrow})              & $\times$                             & \checkmark                          & \checkmark                           & $\times$             & \cite{201567}                                                              \\
\cite{hsu2018scout}   & Search        & Pairwise Modelling (\textit{Scout})                & \checkmark                           & \checkmark                          & \checkmark                           & $\times$             & RS, CD, \cite{201567}                                                      \\
\cite{Micky}          & Search        & Multi-armed Bandits (\textit{Micky})                  & $\times$                             & \checkmark                          & $\times$                             & $\times$             & \cite{201567}                                                              \\
Ours                  & Search        & RBFOpt, Hyperopt, SMAC, CloudBandit   & $\times$                             & \checkmark                          & $\times$                             & \checkmark           & RS, \cite{194946}, \cite{Yadwadkar}, \cite{201567}, \cite{Bilal} \\ \hline
\end{tabular}
\end{table*}

\subsection{Predictive Methods}

Given a target workload, predictive methods use a statistical model to predict how the workload will perform on a set of cloud configurations.
Using these predictions, the cloud configurations can be ranked and the optimal configuration estimated.
Methods vary in terms of what type of model is used, as well as with what kind of data the model is trained. 

In \cite{194946}, the authors considered the problem of finding the optimal type and number of nodes for distributed analytics workloads using a single cloud provider.
The proposed method, Ernest, trains a linear performance model for each node type, using data collected from online evaluations of the target workload.
To minimize the cost of the online evaluations, the workload is evaluated using a small subsample of the workload input data.
The trained model is then used to estimate the performance of the workload using the full input data.


Rather than rely on online evaluations of the workloads, other methods opt to train a predictive model using \textit{offline} data collected from benchmarking or profiling of historical workloads. 
For instance, \cite{MARIANI2018618} proposed an offline approach targeting the configuration of distributed scientific computing workloads. 
To generate the offline dataset, several training workloads were evaluated on a set of cloud configurations.
Using this data, a random forest (RF) model was trained to predict the cloud performance using features related to both the configuration and workload characteristics. 
In the online phase, the same set of features are generated for the target workload and the RF estimates are used to identify the best configuration. 
RF models were also employed in \cite{Yadwadkar} to predict the performance of single-node workloads, such as video compression and cloud data stores. 
The proposed method, PARIS, trains an RF model using both historical data (collected offline across two cloud providers) and online evaluations of the target workloads on a small set of reference configurations. 
The features used to train the model comprise configuration-specific values as well as a \textit{fingerprint} obtained by profiling the target workload on the reference configurations. 
In \cite{216005} a similar approach, combining offline data with online evaluations of the target workload, was proposed to configure single-node analytics workloads.
Rather than using an RF model, latent factor collaborative filtering was used to rank a set of cloud configurations 
from a single cloud provider.



\subsection{Search-based Methods}
\label{sec:soa_search}

While predictive methods usually incur a relatively small online search cost, they suffer from two main drawbacks. 
Firstly, despite large amounts of offline data available, any predictive model is inherently noisy -- for example, PARIS \cite{Yadwadkar} achieves the relative RMSE between $15\%$ and $65\%$.
As a result, inaccurate predictions can lead to a sub-optimal configuration being chosen.
Secondly, predictive methods require a significant amount of offline data, which may be unavailable or time consuming and expensive to collect. 

Thus, an orthogonal research direction has explored search-based methods that iteratively evaluate the target workload using a sequence of different configurations.
The length of this sequence (i.e., the total number of configurations evaluated) is referred to as the \textit{search budget}.
Similarly to predictive approaches, search-based methods can use both offline data and online evaluations.
However, search-based methods additionally use feedback from all previously evaluated configurations in order to better select which configurations should be evaluated next.
Thanks to the additional feedback, search-based methods are more likely to find the true optimal configuration and typically do not require the offline phase. 
However, they may incur a relatively large online search runtime and cost. 
Thus, such methods are most appropriate for frequently recurring workloads, such as daily log parsing or re-training of ML models, for which the cost of the configuration search (which should be run relatively infrequently) can be effectively amortized. 
Recurring workloads are an important use-case: in one production cluster, recurring workloads were found to account for over 40\% of jobs \cite{10.1145/2168836.2168847}.

The simplest search-based method, and an important baseline, is exhaustive search (i.e., trying all possible configurations across all cloud providers). 
While this approach is guaranteed to find the optimal configuration, it can be extremely time consuming and expensive.
Other simple baselines such as random search (RS) and coordinate descent (CD), may be less expensive than exhaustive search, but may fail to find the optimal configuration for a small search budget.
More efficient search-based methods are desirable so that their cost can be more effectively amortized, allowing them to be run more frequently to adapt to changes in the workload characteristics. 

In \cite{201567} the authors pose the cloud configuration problem as an optimization problem in which one would like minimize the cost of running a workload subject to a runtime constraint.
To solve this, they propose an online system, CherryPick, that uses Bayesian Optimization (BO) with a Gaussian process (GP) surrogate, a Matern5/2 kernel and an Expected Improvement (EI) acquisition function that has been modified to account for the runtime constraint. 
Experiments are performed for 5 different distributed analytics workloads for 66 cloud configurations 
from a single cloud provider.
It was demonstrated that CherryPick has a high probability of finding the optimal configuration, with a lower online search cost and time relative to RS, CD and Ernest.

BO applied to cloud configuration was further investigated in \cite{Bilal}.
The authors frame cloud configuration as an unconstrained optimization problem in which the goal is to minimize either the runtime or cost of executing a workload in the cloud.
Different BO surrogate functions were considered in addition to GP: RF, gradient-boosted regression trees (GBRT) and extra-trees (ET). 
The authors perform experiments for 12 distributed analytics and ML workloads across a maximum of 140 cloud configurations from a single provider.
The various BO flavours are compared with other search-based solutions, such as stochastic hill climbing, simulated annealing and Tree-structured Parzen Estimator.
Their conclusions were threefold: (a) in general, BO outperforms the other methods considered, (b) BO with GP and lower confidence bound (LCB) is the best choice when optimizing for cost, and (c) BO with GBRT or RF and probability of improvement (PI) is the best choice when optimizing for runtime.

Further improvements to BO-based methods were proposed in \cite{8416333}.
The authors propose augmenting BO with low-level information captured from the workloads during execution, such as CPU utilization, or working memory size. 
Additionally, the authors propose using the ET surrogate together with a different acquisition function: prediction delta. 
Experimental results were presented for 107 workloads across 88 different configurations from a single cloud provider.
For 43\% of workloads, the augmented form of BO was found to outperform naive GP-based BO, in terms of performance as well as search cost.

A number of works have also explored search-based methods other than BO.
In \cite{hsu2018scout} an approach was proposed, Scout, that leverages offline data to guide a online search algorithm based on pairwise-modelling.
Scout was evaluated on 3 distributed big data analytics workloads across 69 configurations from a single cloud provider and was shown to outperform CherryPick, RS and CD in terms of execution time, deployment cost and search cost.
The dataset from \cite{hsu2018scout} was made publicly available, and was used in \cite{Micky} to evaluate a search-based method, Micky, based on multi-armed bandits. 
Unlike all of the works mentioned so far, Micky tries to solve a different problem: finding the best cloud configuration for a \textit{group} of workloads.
It is shown that when the size of this group is large, Micky can find a satisfactory solution with a lower search cost than CherryPick.

A summary comparison of all aforementioned works is provided in Table \ref{tab:soa}. We compare predictive and search-based methods in terms of algorithms used, whether they involve offline or online evaluations, whether they leverage low-level metrics captured from the workload, whether multi-cloud environments were considered and finally, what baselines were used in their experiments.

\subsection{Related Problems}

A number of works have studied a similar optimization problem in the context of cloud scheduling and other fields, both for single-cloud \cite{topcuoglu2002performance} and multi-cloud environments \cite{van2010cost}.
In this line of research, a workload is typically modelled as a directed acyclic graph, in which the nodes represent individual tasks, and the edges represent dependencies between tasks.
Each task can be assigned to a different cloud, and assigned a different configuration, and the goal is to minimize the \textit{makespan} of the workload (i.e., end-to-end runtime), possibly under various constraints.
Unlike the predictive and search-based methods reviewed above, which evaluate the performance of workloads via real cloud measurements, these methods assume that the runtime of each task on a given configuration is either known a-priori or is given by some simple theoretical model based on the compute capability of the node in GFLOPS \cite{durillo2014multi}.
This assumption gives rise to a variety of different solutions: graph-based heuristics \cite{topcuoglu2002performance, zheng2013budget, durillo2014multi}, genetic algorithms \cite{liu2017deadline, shi2020location}, integer linear programming \cite{van2010cost}, and multiple criteria decision making \cite{9284492}.
To the best of our knowledge, none of these methods have been evaluated in a real cloud environment, but on simulation data. 


\newpage
\section{Search-based Multi-Cloud Configuration}
 \label{section:method}

\subsection{Problem Statement}

The runtime/cost of a workload in the cloud depends on many factors, such as the number of nodes used, the vCPU count, the amount of memory allocated to each node, the manufacturer/generation of the backend CPU, the network on which the nodes communicate and the region in which the nodes are deployed.
While some of these parameters (e.g., vCPU count and amount of memory) are, on the surface, common across different cloud providers, most providers do not give the option to set these parameters freely.
Instead, one must select from a set of available VM types, parameterized into categories like \textit{family}, \textit{type} or \textit{size}, that define which CPU will be used, how many vCPUs and how much memory will be assigned to the node, as well as what network interfaces are used. 
These categories are significantly different across cloud providers, making it difficult to construct a multi-cloud optimization problem over parameters that are common across all cloud providers (e.g., vCPU count) without introducing complex constraints.

Instead, it is much easier to view the multi-cloud configuration problem as an optimization over a hierarchical domain, where the domain for each cloud provider comprises a unique set of categorical parameters.
Let $\mathcal{K}=\{1,2,\ldots,K\}$ denote the set of available cloud providers, and let $\mathcal{X}^{(k)} = \mathcal{C}^{(k)}_1 \times \dots \times \mathcal{C}^{(k)}_{s_k}$ denote the set of parameter configurations offered by the $k$-th cloud provider, where $\mathcal{C}^{(k)}_{i}, i=1, \dots, s_k$ denotes the set of values of the $i$-th configuration parameter of $k$-th cloud provider and $s_k$ denotes the number of configuration parameters available for $k$-th cloud provider. Additionally, let $\mathcal{N}\subset\mathbb{Z}_{\geq 1}$ denote the set of valid sizes of the Kubernetes cluster in terms of number of nodes.
Then, the multi-cloud configuration problem can be posed as the following hierarchical optimization problem:
\begin{equation}
	\min_{k\in\mathcal{K}}\left[\min_{n\in\mathcal{N}, \vec{x}\in\mathcal{X}^{(k)}} f_k(n, \vec{x})\right],\label{eq:mcc-opt}
\end{equation}
where objective function $f_k(n, \vec{x})\in\mathbb{R}^+$ returns the execution time (or cost) of the target workload on a Kubernetes cluster deployed on the $k$-th cloud provider, comprising $n$ nodes of type $\vec{x}$.
As runtime and cost of running the workload in cloud with a specific configuration is not deterministic, function $f_k$ can return the target value obtained in a single measurement or use any chosen metric based on multiple measurements, such as the mean or the 90th percentile.

In this work, we are interested in developing search-based methods that try to solve problem \eqref{eq:mcc-opt}. 
The search for the best configuration will be constrained by a search budget $B\in\mathbb{Z}_{\geq 1}$, specified in terms of the maximum number of evaluations of the objective functions $f_k$. 

\subsection{Adapting State-of-the-Art to Multi-Cloud}
\label{section:adaptations}

\begin{figure*}[t]
\centering
\begin{subfigure}{.27\textwidth}
  \centering
  \includegraphics[width=\linewidth]{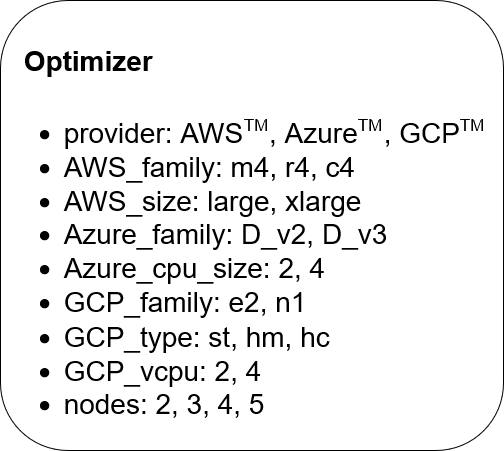}
  \caption{}
  \label{subfig:x1}  
\end{subfigure}\hfill
\begin{subfigure}{.7\textwidth}
  \centering
  \includegraphics[width=\linewidth]{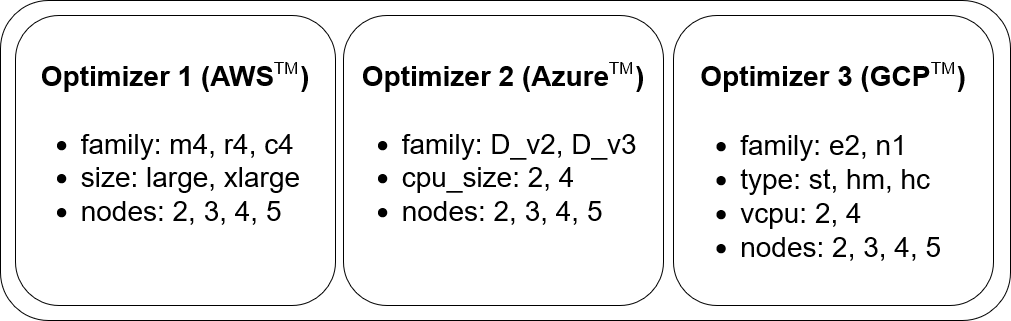}
  \caption{}
  \label{subfig:x3}
\end{subfigure}
\caption{Domains resulting from two adaptations of state-of-the-art optimizers to the multi-cloud setting: a) Flattened domain (single optimizer instance), and b) Hierarchical domain ($K$ independent optimizer instances).}
\label{fig:x1_vs_x3}
\end{figure*}

As discussed in Section \ref{sec:soa_search}, the majority of related work was focused on applying black-box optimization methods, in particular BO \cite{201567, Bilal, 8416333}, to the single-cloud configuration problem. 
It is thus natural to ask how such methods might be adapted to solve the multi-cloud configuration problem defined in \eqref{eq:mcc-opt}.
Below we consider two such approaches, which are illustrated visually in Figure \ref{fig:x1_vs_x3}.

\subsubsection{Flattening the domain}
\label{sec:flattened}
The main challenge when applying off-the-shelf BO to \eqref{eq:mcc-opt} is that the domain of the inner minimization depends on the choice of cloud provider in the outer minimization. 
One straightforward approach is to effectively \textit{flatten} the optimization domain by defining a new objective function that includes the cloud provider selection as an argument:
\begin{equation}
	f\left(k, n, \vec{x}^{(1)}, \vec{x}^{(2)}, \ldots, \vec{x}^{(K)}\right) = f_k(n, \vec{x}^{(k)}),\label{eq:obj-flat}
\end{equation}
and solve the equivalent optimization problem:
\begin{equation}
	\min_{\substack{k\in\mathcal{K}, n\in\mathcal{N} ;\\ \vec{x}^{(1)}\in\mathcal{X}^{(1)}, \ldots, \vec{x}^{(K)}\in\mathcal{X}^{(K)}}}  f\left(k, n, \vec{x}^{(1)}, \ldots, \vec{x}^{(K)}\right).\label{eq:mcc-opt-flat}
\end{equation}
This approach is illustrated visually in Figure \ref{subfig:x1}.
While optimization problem \eqref{eq:mcc-opt-flat} can be readily solved using standard BO software packages, it has two key downsides. 
Firstly, the dimensionality of the flattened domain $\mathcal{K}\times\mathcal{N}\times\mathcal{X}^{(1)}\times\cdots\times\mathcal{X}^{(K)}$ becomes large as the number of cloud providers increases.
It is generally understood that BO does not perform well in this limit \cite{wang2013bayesian}, often being beaten by simple baselines like random search.
Secondly, for a given value of $k$, the choice of $\vec{x}^{(k')}$ for $k'\neq k$ will have no effect on the objective function \eqref{eq:obj-flat} and the optimizer may well waste valuable evaluations trying to understand these non-existent relationships.

\subsubsection{Independent optimizers}
\label{sec:independent_optimizers}
An alternative, perhaps even simpler, way to adapt off-the-shelf BO to \eqref{eq:mcc-opt} is simply to use $K$ independent optimizers, one for each cloud provider. 
Thus, the $k$-th optimizer will solve the inner optimization problem:
\begin{equation}
	\min_{n\in\mathcal{N}, \vec{x}\in\mathcal{X}^{(k)}} f_k(n, \vec{x}) \label{eq:mcc-opt-separate}.
\end{equation}
This approach is illustrated visually in Figure \ref{subfig:x3}.
In order to make a fair comparison against the flattened approach, it is necessary that the search budget be split equally amongst each of the $K$ optimizers.
Specifically, if the single optimizer over the flattened domain is given budget $B$, then each of the $K$ independent optimizers should be given budget $B/K$.
The approach has the advantage that it acknowledges the hierarchy that is present in the problem, albeit in the most naive way, and, as a result, each optimizer is operating over a significantly smaller domain. 
However, it has the disadvantage that all cloud providers will be assigned equal budget, and thus during the search it cannot learn to discard cloud providers that do not look promising. 
Additionally, the parameter corresponding to the number of nodes is optimized independently for each cloud provider, when it is in fact a parameter that is common to all.

\subsection{Applying Methods from AutoML}

In \cite{Bilal} the authors make an analogy between cloud configuration and hyperparameter tuning in ML.
They point out important differences between these two problems. 
Firstly, cloud configurations consist mostly of integer or categorical variables, whereas hyperparameters in ML are often continuous.
Secondly, cloud configurations are typically complex and expensive to evaluate (involving spinning up a cluster of nodes), whereas hyperparameter configurations may be easily evaluated on a single node.
Despite these differences, the authors found that methods that work well for hyperparameter tuning (such as BO with RF or GBRT) are also highly effective for solving the cloud configuration problem.

The key insight of this part of our work is to go one step further and notice that, due to its hierarchical nature, the selection-configuration problem defined in (\ref{eq:mcc-opt}) bears a striking resemblance to the Combined Algorithm Selection and Hyperparameter Optimization (CASH) problem at the heart of AutoML.
Namely, in AutoML one must \textit{select} which ML model to use (e.g., neural networks vs. gradient-boosted decision trees), and decide how to \textit{configure} each model (e.g., how each hyperparameter should be chosen). 
Having established this analogy, it makes sense to ask the question: can existing methods from AutoML be applied to the multi-cloud configuration problem defined in (\ref{eq:mcc-opt})? 


HyperOpt and SMAC are both examples of sequential model-based optimization (SMBO) algorithms based on BO, commonly used in AutoML.
SMAC, introduced in \cite{10.1007/978-3-642-25566-3_40} and recently extended in \cite{lindauer2021smac3}, implements a form of BO with an RF surrogate.
HyperOpt was introduced in \cite{pmlr-v28-bergstra13} and uses the Tree-Structured Parzen Estimator (TPE) as a surrogate.
Both optimizers are able to take into account the natural hierarchy present in selection-configuration problems.
SMAC achieves this by modeling the hierarchical domains using an RF, whereas HyperOpt models it as a graph-structured generative process, which sets valid configuration parameters sequentially.
They are both actively maintained GitHub projects and provide APIs that can be used to solve arbitrary selection-configuration problems (rather than being specific to AutoML).
Therefore, they can be easily applied to multi-cloud configuration.

Another solution used to solve the CASH optimization problem and capable of utilizing the hierarchical structure of the domain is Rising Bandits (RB), a best-arm identification approach proposed in \cite{Li_Jiang_Gao_Shao_Zhang_Cui_2020}.
RB defines multiple arms, each corresponding to a different ML model.
Pulling an arm corresponds to running a fixed number of iterations of BO to find the best hyperparameter setting for the corresponding ML model.
The goal of the optimization is to identify the best arm, which maximizes the reward (given by the best validation accuracy found by the optimizer).
To have an efficient algorithm, the goal is to achieve this whilst minimizing the total number of arm pulls.
To that end, RB makes a theoretical assumption about the way the model's accuracy converges as a function of BO iterations.
Specifically, RB assumes that after a certain number of pulls all arms will reach a point of \textit{diminishing returns}.
When this point is reached, the authors derive a theoretical criterion for when the arm is guaranteed to never outperform another available arm, at which point it is eliminated.
Thanks to arm eliminations, more budget can be devoted to more promising ML models.

Adapting a best-arm identification scheme such as RB to the multi-cloud configuration use case is relatively straightforward.
Firstly, each arm is associated with a different cloud provider.
Each arm pull corresponds to running an optimizer searching for the best cloud configuration and number of nodes for the respective cloud provider.
Rather than trying to maximize accuracy, in this context the goal of RB is to minimize runtime or cost.
Unfortunately, the theoretical assumption made by RB is not guaranteed to hold in a multi-cloud configuration scenario.
As a result, RB may incorrectly choose the arm to be eliminated.
Despite this, we include RB in our experimental work.

\subsection{A Novel Bandit-based Approach: CloudBandit}
\label{subsection:SE}

RB is promising in the AutoML context but its assumptions cannot be translated to the multi-cloud scenario.
Therefore, we develop a new bandit-based algorithm, \namealgo{} (\shortalgo{}), specifically designed for the multi-cloud configuration use case. 
\shortalgo{} is different to RB in three important aspects.
Firstly, each arm pull in \shortalgo{} corresponds to running an arbitrary black-box optimizer (BBO) rather than using only BO.
Secondly, the search budget for each arm grows by a multiplicative factor as subsequent arms are eliminated.
Thanks to this, more promising cloud providers are explored exponentially more than those that are eliminated.
Finally, \shortalgo{} uses a simple empirical method for eliminating arms, rather than relying on any theoretical assumptions.

We describe \shortalgo{} in detail in Algorithm \ref{alg:bandit}.
The algorithm maintains an \textit{active set} of arms $A$, which is initially equal to the complete set of all arms (i.e., all cloud providers $\mathcal{K}$).
Pulling an arm corresponds to running a single iteration of an arbitrary BBO searching for the best configuration and number of nodes for the corresponding cloud provider.
Let $p_{k, t} \in \mathcal{X}^{(k)} \times \mathcal{N}$ denote the best configuration and number of nodes found by pulling the $k$-th arm $t$ times and let loss $L_{k, t}=f_k(p_{k, t})$ denote the runtime or cost for this (configuration, number of nodes) pair.
\shortalgo{} performs a number of rounds equal to the number of providers $K$.
In the $m$-th round (for $m=1\ldots,K$), each arm in the active set $A$ is pulled $b_m$ times, i.e. we run $b_m$ iterations of an arbitrary BBO algorithm to find the best configuration and number of nodes for the corresponding cloud provider.
Therefore, in $m$-th round \shortalgo{} finds the best (configuration, number of nodes) pair $p_{k, \hat{b}+b_m}$ and corresponding loss $L_{k, \hat{b}+b_m}$ for each active arm $k \in A$, where $\hat{b} = \sum_{i=1}^{m-1} b_i$ denotes the total number of pulls performed on each active arm in past rounds.
Next, the arm in the active set with the highest loss $L_{k, \hat{b}+b_m}$ is identified and eliminated from the active set $A$.
Before proceeding to the next round, the search budget is increased by a multiplicative factor $\eta$.
At the end of the $K$-th round, the best identified (configuration, number of nodes) pair is returned, for the sole remaining provider.

\begin{algorithm}[ht]
\begin{algorithmic}[1]
\State Initialize set of arms (cloud providers): $A\!\!=\!\!\mathcal{K}\!\!=\!\!\{ 1, \ldots, K \}$ and $\hat{b}=0$. Choose component BBO $\mathcal{O}$. 
\State Set initial budget $b_1$ and budget growth factor $\eta$.
\For {round $m=1,\ldots,K$}
	\For {cloud provider $k \in A$}
		\State Run $b_m$ iterations of optimizer $\mathcal{O}$ for provider $k$.
		\State Receive best (configuration, nodes) pair $p_{k,\hat{b}+b_m}$ and corresponding loss: $L_{k,\hat{b}+b_m}=f_k(p_{k,\hat{b}+b_m})$.
	\EndFor
	\State Identify the worst arm and eliminate it from the active set:  $A = A \setminus \{\argmax_{k\in A} L_{i, \hat{b}+b_m}\}$.
	\State Increment $\hat{b}=\hat{b}+b_m$ and set budget for next round: $b_{m+1} = \eta \cdot b_m$.
\EndFor
\State Output best provider $k^*=\argmin_{k\in A} L_{i, \hat{b}}$ and its best (configuration, nodes) pair $p_{k^*, \hat{b}}$.
\end{algorithmic}
\caption{\namealgo{}}
\label{alg:bandit}
\end{algorithm}

The algorithm has two hyperparameters: the initial budget $b_1$ and the growth factor $\eta$.
The total search budget of \shortalgo{} can be expressed in terms of these two hyperparameters: $B=\sum_{m=1}^K (K\!-\!m\!+\!1) b_1 \eta^{m-1}$.
By using relatively small $b_1$ and setting $\eta>1$, it is hoped that the algorithm can eliminate slow (or expensive) cloud providers very quickly and devote exponentially more budget to the more promising cloud providers.

\newpage
\section{Experimental Results}
\label{section:experiments}

\begin{table}[thbp]
\caption{Details of the optimization tasks and cloud configuration parameters used in the dataset.}
\label{tab:dataset}
\begin{tabular}{@{}lll@{}}
	\hline
	Dask tasks &  \multicolumn{2}{l}{\makecell[l]{kmeans, linear regression, logistic regression, naive bayes, \\poisson regression, polynomial features, spectral clustering, \\quantile transformer, standard scaler, xgboost}} \\ \hline
	Datasets & \multicolumn{2}{l}{buzz in social media \cite{buzz}, credit card \cite{creditcard}, santander \cite{santander}} \\ \hline
	Targets & \multicolumn{2}{l}{cost, runtime} \\ \hline
	\multirow{3}{*}{\makecell[l]{Cloud\\configuration}} & AWS\textsuperscript{\texttrademark} & \makecell[l]{family: m4, r4, c4\\size: large, xlarge} \\ \cline{2-3}
													& Azure\textsuperscript{\texttrademark} & \makecell[l]{family: D\_v2, D\_v3\\cpu\_size: 2, 4} \\ \cline{2-3}
													& GCP\textsuperscript{\texttrademark} & \makecell[l]{family: e2, n1\\type: standard, highmem, highcpu\\vcpu: 2, 4} \\ \hline 
	Nodes & \multicolumn{2}{l}{2, 3, 4, 5} \\ \hline
\end{tabular}
\end{table}

\subsection{Building an Offline Benchmark Dataset}

Comparing search-based methods in an online setting requires time-consuming and expensive evaluations each and every time an algorithm is run. 
Therefore, in order to compare the different methods described in Section \ref{section:method} in a time and cost-effective way, it was necessary to first construct an offline dataset that can be used for benchmarking.
This dataset could then be used to evaluate all of the different methods, including many repetitions with different random seeds, by \textit{simulating} the behaviour of the algorithms. 
Specifically, when the algorithm needs to evaluate a workload using given configuration on a given provider, rather than going and creating a Kubernetes cluster and deploying the workload, we can simply read the runtime or cost from the offline dataset that was previously collected. 
While a similar dataset has previously been published for the single-cloud configuration problem \cite{hsu2018scout}, there exists no public data regarding multi-cloud configuration\footnote{We intend to make our dataset public upon publication of this paper, and have already obtained the necessary internal approvals required for doing so.}.

\textbf{Workloads and optimization tasks.} 
When constructing our dataset, we considered a variety of distributed data analytics tasks implemented using the Dask framework \cite{dask2016}.
Dask was selected since it has a clean integration with Kubernetes, and supports a wide variety of tasks including ML training, as well as data pre-processing. 
We tried to select a set of tasks with significantly different characteristics.
For example, distributed training of gradient-boosted decision trees (XGBoost \cite{xgboost}) involves plenty of branching logic and complex communication patterns, whereas distributed clustering tasks like k-means are typically compute-bound with minimal communication \cite{10.1145/380995.381010}. 
Each selected Dask task was used to generate \textit{optimization tasks}, each comprising: (a) a workload defined as a (Dask task, input dataset) pair, and (b) an optimization target which can be either runtime or cost.
We collected data for $30$ workloads ($10$ Dask tasks running on $3$ input datasets) and $2$ optimization targets, leading to a total of $60$ optimization tasks. 
Details regarding all optimization tasks can be found in Table \ref{tab:dataset}.

\textbf{Cloud provider and configuration space.} 
Each of the $30$ workloads was then deployed on a variety of different configurations across three different cloud providers: Amazon Web Services\textsuperscript{\texttrademark} (AWS\textsuperscript{\texttrademark}), Microsoft Azure\textsuperscript{\texttrademark} and Google Cloud Platform\textsuperscript{\texttrademark} (GCP\textsuperscript{\texttrademark}).
For each cloud provider, we selected the region that was closest to our research lab geographically.
For AWS\textsuperscript{\texttrademark}, we generated 24 different configurations by varying the cluster size, together with the \texttt{family} and \texttt{size} options. 
For Azure\textsuperscript{\texttrademark}, we varied \texttt{family}, \texttt{cpu\_size} together with the cluster size to generate $16$ different configurations.
Finally, for GCP\textsuperscript{\texttrademark} we generated 48 different configurations by varying the cluster size, \texttt{family}, \texttt{type} and \texttt{vcpu}. 
This resulted in a total of 88 different multi-cloud configurations, specific details of which are also provided in Table \ref{tab:dataset}.
This was the largest configuration space we could evaluate given our research budget constraints.
For each workload on each configuration, we measured the total execution time.
Additionally, we estimated the execution cost in the same manner as \cite{201567, 194946} and \cite{Yadwadkar} by multiplying the runtime by the listed price-per-hour for each type of node, together with a factor equal to the number of nodes.
While this is an imperfect estimate, and will not take into account additional data transfer costs, more precise estimates are difficult to obtain since cloud billing typically occurs monthly in an aggregate manner.

\subsection{Experimental Methodology and Implementation Details}

Using the offline dataset we have collected, we evaluated various predictive and search-based methods from the cloud configuration literature adapted to the multi-cloud setting, search-based methods from the AutoML literature applied to multi-cloud configuration, as well as the new algorithm we have proposed in this paper (\namealgo{}).
For each method, experiments were performed for all 30 workloads, for both time and cost optimization targets.
For the search-based methods, we vary the search budget $B=11,22,\ldots,88$ and perform 50 repetitions using different random seeds. 
Methods are compared in terms of their regret: the relative distance to the true minimum, averaged over all seeds and workloads.
Below, we provide details regarding how each of the methods was implemented.

\textbf{Linear Predictor.} We implemented a purely-online predictive method that builds a linear model for each workload and each cloud configuration using the features defined by Ernest \cite{194946}. 
Unlike Ernest, which trains the model using online evaluations taken using a small subset of the workload input dataset, we train the linear model using online evaluations of the workload using the full input dataset following a leave-one-out approach.
Specifically, for a given workload, cloud provider and a provider-specific configuration, we estimate the runtime/cost for a given cluster size $n\in\mathcal{N}$ by training the linear model using evaluations on all clusters of size $n'\in\mathcal{N}\setminus n$.
Relative to Ernest, this approach has a larger online cost, but should provide a strictly best-case scenario in terms of predictive performance. 

\textbf{Random Forest Predictor.} Following a similar approach to PARIS \cite{Yadwadkar}, we build a set of RF models, one for each cloud provider, to predict the runtime/cost of a given workload on a given configuration. 
Each model is constructed over features related to the configuration, as well as \textit{fingerprint} features corresponding to the performance of the workload on $2$ reference configurations. 
Unlike \cite{Yadwadkar}, our fingerprints simply contain the execution time (or cost) of the workload, rather than low-level information captured from the cluster, as our work is focused on a scenario where the workload is a black box.
To make predictions for a target workload $w$, the model is first trained using feature vectors corresponding to all workloads $w'\neq w$ in the offline dataset. 
Then, in order to construct the fingerprint for the target workload so that a prediction can be made, it is necessary to perform online evaluations using a total of $6$ reference configurations, $2$ for each cloud provider.

\textbf{Random Search (RS).} We implemented the simplest search-based baseline. 
For a given budget $B$, we select $B$ configurations at random (with replacement, across all cloud providers) and select the best one.

\textbf{Bayesian Optimization (BO).} We used BO tools provided by \texttt{scikit-optimize} \cite{skopt} to implement the scheme proposed in CherryPick \cite{201567}, and the schemes suggested by Bilal et al. \cite{Bilal}.
For CherryPick, we thus use a GP surrogate, a Matern5/2 kernel and the EI acquisition function. 
For the schemes from \cite{Bilal}, we use a GP surrogate and LCB acquisition function when optimizing for cost, and an RF surrogate and PI acquisition function when optimizing for time. 
In order to adapt the BO methods to the multi-cloud scenario, we consider the flattened method defined in Section \ref{sec:flattened}, as well as the approach using independent optimizers, as defined in Section \ref{sec:independent_optimizers}.


\textbf{SMAC \cite{lindauer2021smac3} and HyperOpt \cite{pmlr-v28-bergstra13}.} We applied both SMAC and HyperOpt using the code provided on GitHub, in both cases taking care to make sure that the conditional dependencies resulting from the hierarchical nature of the domain are specified correctly. 

\textbf{Rising Bandits (RB). \cite{Li_Jiang_Gao_Shao_Zhang_Cui_2020}} We implemented Rising Bandits ourselves, as no public code was available.
To that end, for the component optimizer we use the BO implementation provided by \texttt{scikit-optimize} \cite{skopt} with default settings (GP surrogate and gp-hedge acquisition function), as no details regarding the used BO method were provided.
We implement lower and upper confidence bounds of each arm's runtime or cost which guide arm elimination analogously to the author's confidence bounds of the ML model's accuracy.

\textbf{CloudBandit (CB).} We set CloudBandit's hyperparameter $\eta=2$ and use multiple values of $b_1$ to vary the total search budget $B$.
We consider $2$ component BBOs: CherryPick and RBFOpt \cite{costa2018rbfopt}.
RBFOpt is a BBO algorithm based on the Radial Basis Function method, originally proposed in \cite{gutmann2001radial}, and is considered in our experiments because it has been shown to outperform BO in a wide variety of scenarios \cite{nannicini2020implementation}.

\subsection{Comparing Methods from Cloud Configuration Literature}

\begin{figure*}[ht]
\centering
\begin{subfigure}{.49\textwidth}
  \centering
  \includegraphics[width=\linewidth]{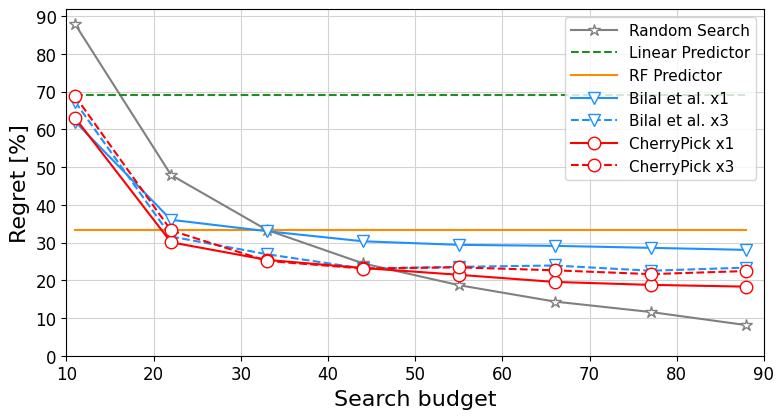}
  \caption{Cost target}
  \label{subfig:results12_cost}
\end{subfigure} \hfill
\begin{subfigure}{.49\textwidth}
  \centering
  \includegraphics[width=\linewidth]{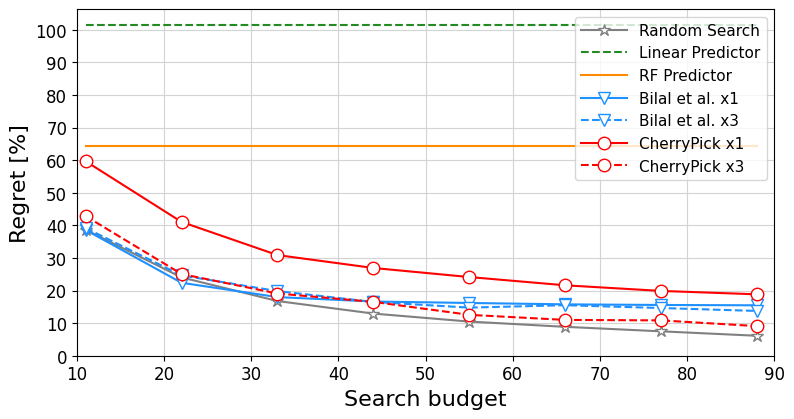}
  \caption{Time target}
  \label{subfig:results12_time}
\end{subfigure}
\caption{Regret of state-of-the-art predictive and search-based cloud solutions adapted for the multi-cloud setting by flattening the domain (`x1') and by using $3$ independent optimizers (`x3'), compared against random search.}
\label{fig:results12}
\end{figure*}

\begin{figure*}[htbp]
\centering
\begin{subfigure}{.49\textwidth}
  \centering
  \includegraphics[width=\linewidth]{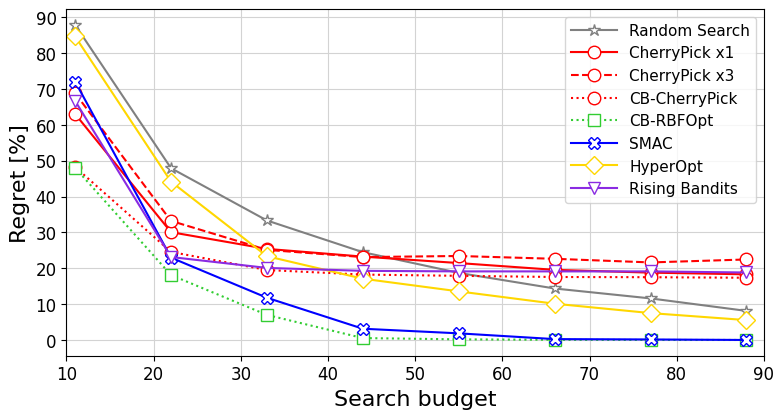}
  \caption{Cost target}
  \label{subfig:results34_cost}
\end{subfigure} \hfill
\begin{subfigure}{.49\textwidth}
  \centering
  \includegraphics[width=\linewidth]{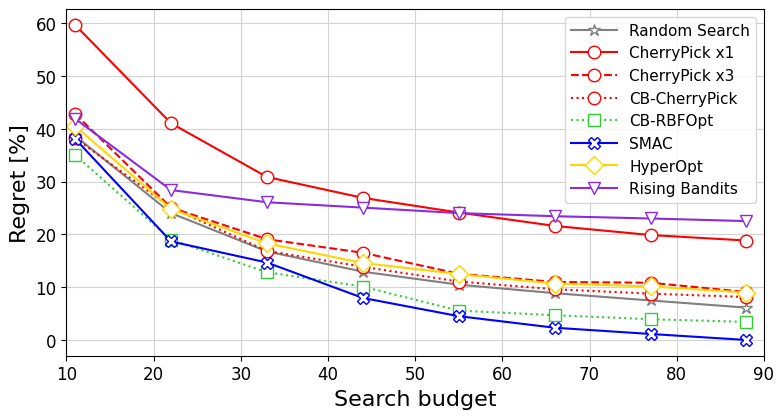}
  \caption{Time target}
  \label{subfig:results34_time}
\end{subfigure}
\caption{Regret of hierarchical AutoML optimizers and the CloudBandit scheme, compared against both adaptations of CherryPick: flattened domain (`x1') and independent optimizers (`x3') and random search.}
\label{fig:results34}
\end{figure*}

In Figure \ref{fig:results12}, we present a comparison between the predictive methods, random search (RS) and BO methods from the cloud configuration literature. 
As there is no well-defined notion of search budget for predictive methods, we present their results as horizontal lines.
For the BO methods, we compare the approach using the flattened domain (denoted `x1'), together with the approach using independent optimizers for each cloud provider (denoted `x3') for a range of search budgets $B$.
From these results, we can draw a number of interesting conclusions:
\begin{itemize}
	\item While the predictive methods, in particular the RF-based approach, identify a relatively good configuration, search-based methods are able to identify configurations with a significantly lower regret, particularly when optimizing for time.
	\item Both adaptations of state-of-the-art BO-based methods (CherryPick and the solution by Bilal et al.) are outperformed by RS in the majority of cases. They beat RS only for the cost optimization target with lower budgets.
	\item While in most cases the approach using independent optimizers (`x3') achieves lower regret than the approach using the flattened domain (`x1'), it is not consistently better. Thus, it seems that neither approach is capable of adequately capturing the structure that is inherent to the multi-cloud configuration domain.
\end{itemize}

\subsection{Comparing Hierarchical Methods}

In Figure \ref{fig:results34} we compare the performance of the three search-based methods from the AutoML literature (SMAC, HyperOpt and RB), against CB with both component optimizers (CB-CherryPick and CB-RBFOpt). Additionally, we include the `x1' and `x3' adaptations of CherryPick, as well as RS, for reference.
From these results, we can draw the following conclusions:

\begin{figure*}[htbp]
\centering
\begin{subfigure}{.49\textwidth}
  \centering
  \includegraphics[width=\linewidth]{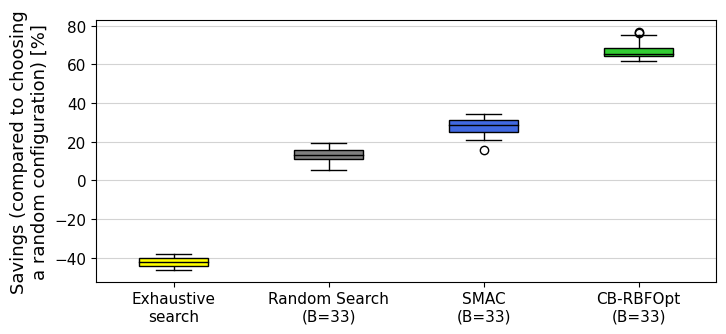}
  \caption{Cost target}
  \label{subfig:savings_cost}  
\end{subfigure} \hfill
\begin{subfigure}{.49\textwidth}
  \centering
  \includegraphics[width=\linewidth]{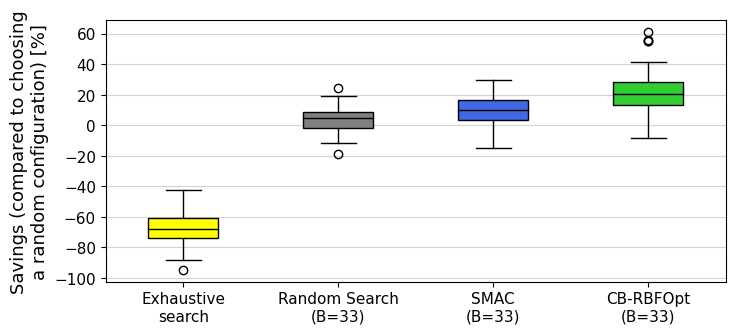}
  \caption{Time target}
  \label{subfig:savings_time}
\end{subfigure}
\caption{Savings of best multi-cloud solutions, exhaustive search and RS compared against choosing a random provider and configuration, for a fixed search budget $B=33$ and fixed $N=64$.}
\label{fig:results4}
\end{figure*}

\begin{itemize}
	\item By explicitly acknowledging the hierarchy that is present in the problem, SMAC is able to outperform almost all methods in terms of regret, for both small and large budgets.
	\item HyperOpt performs well for large budgets, consistently outperforming both adaptations of CherryPick, but it is beaten by both \shortalgo{}-RBFOpt and SMAC. We attribute this to the fact that unlike SMAC, HyperOpt can repeat configurations, which effectively wastes precious evaluations.
	\item RB performs relatively well for small budgets, but is significantly worse than the other AutoML methods for large budgets. We believe this is evidence that the theoretical criterion that is used to eliminate arms does not translate well to the multi-cloud setting.
	\item \shortalgo{}-RBFOpt consistently beats HyperOpt, RB and CherryPick. It also outperforms SMAC for the cost optimization target with small search budgets and is only slightly worse than SMAC for the time optimization target.
	\item Embedding CherryPick inside \shortalgo{} significantly improves its performance, consistently beating both the `x1' and the `x3' adaptations. This, together with excellent results of \shortalgo{}-RBFOpt, demonstrates that \shortalgo{} is capable of identifying the least promising providers early on in the search, and thus it can devote more budget to more promising providers and find the best configurations faster.
	\item SMAC and \shortalgo{}-RBFOpt are the only optimizers that consistently beat RS across all budgets and targets.
\end{itemize}

\subsection{Savings Analysis}

We proceed to compare the best methods identified above in terms of cost and time savings that they can achieve in production.
When a user chooses to use any of the multi-cloud configuration solutions discussed above, they agree to the additional, one-time expense (cost or runtime) carried by the optimization process.
Then they proceed to run the batch workload $N$ times, achieving some savings each time thanks to the optimized cloud deployment.
A natural question is whether the savings incurred during the $N$ production runs of the workload make up for the additional expense of the optimization process itself.
If the optimization expense outweighs the savings, it is more beneficial for the user to simply select a random provider and configuration.

For that reason, we analyze the best algorithms developed in this work in terms of savings compared to choosing the cloud provider and configuration randomly.
We define savings as follows:
$$S = \frac{T\!E_{rand} - T\!E_{opt}}{T\!E_{rand}} = \frac{N\!\cdot\! R_{rand} - (C_{opt} + N\!\cdot\! R_{opt})}{N\!\cdot\! R_{rand}},$$
where $T\!E_{opt}$ and $T\!E_{rand}$ denote the total expense (cost or runtime) incurred by the optimized or random configuration strategy, respectively. 
$C_{opt}$ denotes the one-time expense of the optimization process and $R_{opt}$ and $R_{rand}$ denote the expense of running the workload with the optimized or random cloud configuration, respectively.
$N$ denotes how many workload production runs are performed before the optimizer is used to update the cloud configuration again.

In this study, we present results for the optimizers running once per a fixed number of $N=64$ production runs, all given a fixed search budget $B=33$.
For the analysis we selected the best algorithms from all multi-cloud solutions evaluated previously: SMAC and \shortalgo{}-RBFOpt.
As baselines, we present RS and exhaustive search.
The savings were computed using each algorithm's results averaged over all random seeds, separately for each workload.
The box plots in Figures \ref{subfig:savings_cost} and \ref{subfig:savings_time} present the distributions of savings of the analyzed algorithms across the workloads.
The boxes mark the interquartile range (IQR) of the distribution (i.e., the $25$--$75$ percentile range), with the additional line inside marking the median.
The whiskers spread by at most $1.5$ IQR in both directions.

For the time optimization target, the differences between the algorithms' savings are not as distinct as for the cost optimization target, but the tendencies of how the algorithms compare to one another are the same for both targets.
SMAC provides median savings of approximately $30\%$ for the cost target and $10\%$ for the time target, outperforming RS and exhaustive search in terms of savings for both optimization targets.
\shortalgo{}-RBFOpt achieves significantly higher median savings for both optimization targets: approximately $65\%$ for the cost target and $20\%$ for the time target.
This is the result of CloudBandit identifying and focusing on the best cloud providers early on, thanks to which it reduces the number of evaluations made on the more expensive cloud providers.
For the time optimization target, optimizers sometimes achieve negative savings, but only within the $0$--$25$ percentile range.
Additionally, for the cost optimization target, neither algorithm has negative savings for any workload.
This means that both SMAC and \shortalgo{}-RBFOpt can be considered more profitable than simply selecting a random cloud provider and configuration.
Finally, exhaustive search is dramatically outperformed by all other approaches and achieves strictly negative savings for both optimization targets.
Testing all available configurations is therefore not a feasible approach to multi-cloud configuration.

\section{Conclusion}

In this paper, we have performed an in-depth study of the suitability of different predictive and search-based methods for the multi-cloud configuration problem. 
We have first shown that existing search-based methods from the single-cloud configuration literature, adapted to the multi-cloud optimization problem, outperform state-of-the-art predictive methods. However, they are still inferior to random search in the majority of cases, even when adapting them to take into account the hierarchical structure of the multi-cloud optimization domain. 
Furthermore, we have shown that an AutoML solution (SMAC), which natively leverages this hierarchy, exhibits excellent performance when compared to search-based methods from the cloud configuration literature, as well as random search. 
Lastly, we have introduced a novel search-based method, CloudBandit, specifically designed to solve the multi-cloud problem. The experimental results have demonstrated that CloudBandit has similar performance to SMAC when using RBFOpt as an internal black-box optimizer. CloudBandit has two advantages over SMAC: 1) it performs better for the cost optimization target with small search budgets, and 2) it provides larger cost and time savings in production. 
As future work, we are planning to perform a similar in-depth study, but for another category of workloads, namely ML inference applications. 
We further intend to enhance CloudBandit's performance by using a component BBO that incorporates the workload's low-level metrics, such as CPU usage or I/O pressure.





\bibliographystyle{plain}
\bibliography{citations.bib}


\end{document}